\documentclass[a4paper,final]{appolb}
\usepackage{amsmath,euscript,amssymb}
\usepackage{fancyhdr}
\usepackage[dvips]{graphicx}
\usepackage{axodraw,epsfig}
\def\be{\begin{equation}}
\def\ee{\end{equation}}
\def\bc{\begin{center}}
\def\ec{\end{center}}

\newcommand{\bea}{\begin{eqnarray}}
\newcommand{\eea}{\end{eqnarray}}

\begin{document}

\pagestyle{fancy} \fancyhead{} \fancyhead[RO,LE]{\thepage}
\fancyhead[CE]{\scriptsize{\textsc{B.M. Barbashov, L.A. Glinka, V.N.
Pervushin, S.A. Shuvalov, A.F. Zakharov}}}
 \fancyhead[CO]{\emph{Hamiltonian approach to the Standard Model}} \fancyfoot{}
\renewcommand\headrulewidth{0pt}
\renewcommand\footrulewidth{0pt}

\title{\Huge{On Hamiltonian approach to  Standard Model}}

\author{
Boris M. Barbashov$^{1}$\vspace{3pt}\\
\L ukasz A. Glinka$^{1}$\thanks{E-mail to: \texttt{glinka@theor.jinr.ru}}\vspace{3pt}\\
Victor N. Pervushin$^{1}$\vspace{3pt}\\
Sergey A. Shuvalov$^{2}$\vspace{3pt}\\
Alexandr F. Zakharov$^{1,3}$
\address{$^1$N.N. Bogoliubov Laboratory of Theoretical Physics,\\Joint Institute for Nuclear Research, 141980 Dubna, Russia}
\address{$^2$Russian Peoples Friendship University, 117198 Moscow, Russia}
\address{$^3$Institute for Theoretical and Experimental Physics, 117259 Moscow, Russia}
}

\date{\today}

\maketitle

\begin{abstract}
The vector bosons models including Standard Model (SM) are
investigated in the framework of the Dirac Hamiltonian method with
explicit resolving the Gauss constraints in order to eliminate
variables with zero momenta and negative energy contributions in
accordance with the operator quantization principles. The
Hamiltonian formulation admits the dynamic version of the Higgs
potential, where its constant parameter is replaced by the dynamic
zero Fourier harmonic of the very Higgs field. In this case, the
  zero mode  equation is a new sum-rule that
 predicts mass of Higgs field
 $m_h=\sqrt{6m^2_t-4[2M_W^2+M_Z^2]}=311.6\pm 8.9 \mbox{\rm GeV}
$.
The Hamiltonian formulation leads to static interactions playing
the crucial role in the off-mass-shell
 phenomena of the type of   bound state 
 and a kaon - pion
 transition in the weak nonleptonic decays.
\end{abstract}


\newpage

\tableofcontents

\section{Introduction}

The Hamiltonian  approach to gauge theories was considered as the
mainstream of
  development of
 gauge theories beginning with the pioneer papers by Dirac
\cite{d}, Heisenberg  and Pauli \cite{hp}, and finishing by the
Schwinger quantization of the non-Abelian theory \cite{sch2} (see in
detail \cite{pol,6}).
 They
  postulated the higher
 priority of the quantum principles, in
  particular, in accordance with the uncertainty principle,
 one counted that we cannot quantize ''field variables''
 whose velocities are absent in
the Lagrangian. Therefore, vector field time
 components with negative contributions into energy
 are eliminated, as it was accepted in the Dirac approach to QED
 \cite{d}. This illumination leads to the static interactions
 and instantaneous bound states.

Remember that  the Dirac Hamiltonian approach generalized
  to the non-Abelian theory \cite{sch2,6} and the massive
  vector fields \cite{hpp}
 provides the fundamental operator
 quantization   and correct relativistic
 transformations of states of quantized fields.
 This Hamiltonian approach is
   considered    \cite{f1} as
  the foundation of all heuristic methods of quantization
  of gauge theories, including the Faddeev-Popov (FP) method
  \cite{fp1}  used now for description of
   Standard Model of elementary
   particles \cite{db}. Moreover,
  Schwinger
 {\it ... rejected all Lorentz
 gauge formulations as unsuited to the role of
 providing the fundamental operator
 quantization} (see \cite{sch2}  {p.324}).
 However, a contemporary reader could not find
  the Hamiltonian presentation of
  the Standard Model (SM) because there is
  the opinion \cite{f1} that this presentation
  is completely equivalent to the accepted version of SM
  based on
 the FP  method \cite{db}.

In this paper, the Weinberg--Salam Standard
 Model    is studied
 in the framework of the Dirac Hamiltonian method
 with explicit resolving  the Gauss constraints
 in order to eliminate variables with zero momenta
 and  negative contribution in energy.
 We try to reply the following question.
 What are new physical results that following from the
 Hamiltonian approach to QED and SM?

\section{Hamiltonian approach to QED}

\subsection{Action and reference frame}

 Let us recall the Dirac approach \cite{d} to QED. The theory is given by
 the well known action
 \be\label{1e} S=\int d^4x\bigg\{-\frac{1}{4}F_{\mu\nu}F^{\mu\nu}+
 \bar \psi [i\rlap/\partial -m]\psi+A_\mu j^{\,\mu}\bigg\},
 \ee
where $F_{\mu\nu}=\partial_\mu A_\nu -\partial_\nu A_\mu$ is a
tension, $A_{\mu}$ is a vector potential, $\psi$ is the
 Dirac electron-positron bispinor field and
 $j_{\mu}=e  \bar {\psi}  \gamma_{\mu} \psi$ is the charge
 current and $\rlap/\partial =
\partial^{\mu} \gamma_{\mu}$. This action is invariant with respect to the collection of gauge
transformations
 \bea
 \label{3e1}
 A^{\lambda}_\mu=A_\mu+\partial_\mu\lambda,~~~
\psi^{\lambda}=e^{+\imath e\lambda}\psi.
 \eea
 The action principle used for the action (\ref{1e}) gives the Euler-Lagrange equations of motion - known as the Maxwell equations
 \be\label{vp}\partial_\nu F^{\mu\nu}+j^\mu=0,
 \ee
Physical solutions of the Maxwell equations are obtained in a fixed
{\it inertial reference frame} distinguished by a unit timelike
vector \mbox{$n_{\mu}$}. This vector splits the
 gauge field $A_\mu$ into the timelike $A_0=A_\mu n_{\mu}$
 and spacelike \mbox{$A^{\bot}_\nu=A_\nu - n_{\nu}(A_\mu n_{\mu})$}
 components. Now we rewrite the Maxwell equations by components\bea\label{c1}
 \Delta A_0-\partial_0\partial_{k}A_k &=&j_{0},\\\label{jc1}
\Box A_k-\partial_k[\partial_{0}A_0-\partial_iA_i ]&=&-j_{k}.
  \eea
The field component $A_0$
  cannot be a {\it degree of freedom}
   because its canonical conjugate momentum vanishes.
The Gauss constraints (\ref{c1}) have the solution: \bea\label{c2}
  A_0+\partial_0\Lambda=-\frac{1}{4\pi}\int d^3y\frac{j_0(x_0,y_k)}{|\mathbf{x}-\mathbf{y}|},
  \eea
 where
  \bea\label{lc2}
  \Lambda=-\frac{1}{\Delta}\partial_{k}A_{k}=\frac{1}{4\pi}\int{d^3y}\frac{\partial_{k}A_{k}}{|\mathbf{x}-\mathbf{y}|}
  \eea
  is a longitudinal component.
  The result (\ref{c2}) is treated as the {\it Coulomb potential field} leading to
 the {\it static} interaction.

\subsection{Elimination of time component}
 Dirac \cite{d} proposed to eliminating of the time component by the
 substitution of the manifest resolution of the Gauss constraints given by
(\ref{c2}) into the
 initial action (\ref{1e}). This substitution - known as reduction
 procedure -
  allows us to eliminate nonphysical pure gauge degrees of
  freedom \cite{gpk}. After this step the action (\ref{1e}) takes
the form \bea\label{2.3-1}S=\int d^4x
\left\{\frac{1}{2}(\partial_{\mu}A^{\rm T}_k)^2
 \!+\!
 \bar \psi
 [i\rlap/\partial\!-\!m]\psi\!-\!j_0\partial_0 \Lambda\!-\!
 A^{\rm T}_kj_{k}\!+\!\frac{1}{2}
 j_0\frac{1}{\triangle}j_0\right\},
 \eea
 where
 \be\label{lc1}
 A^{\rm T}_k=\left(\delta_{ij}-
\frac{\partial_i\partial_j}{\triangle}\right)A_j.
 \ee
This substitution leaves the longitudinal component
  $\Lambda$
 given by Eq. (\ref{lc2}) without any kinetic term.

 There are two possibilities. The first one is to treat
 $\Lambda$ as the Lagrange factor that leads to
 the conservation law (\ref{vp}).  In this approach, the longitudinal component
is treated
 as an independent variable. This treatment violate gauge invariance because
 this component is gauge-variant one and it  be can not measurable.
 Moreover, the time derivative of the longitudinal  component in Eq. (\ref{c2}) looks like
 a physical source of the Coulomb potential. By these reasons we will not consider this
approach in this paper.

 In the second possibility a measurable
 potential stress is identified with the gauge-invariant quantity  (\ref{c2})
 \be\label{2-3-4}
 A_0^{\rm R}=A_0-\frac{\partial_0\partial_k }{\triangle}A_k~,
 \ee
 This approach  is consistent with    the
 principle of gauge invariance
  that identifies
 observables  with gauge-invariant quantities.
 Therefore according to the gauge-invariance
   the longitudinal
 component should be eliminated from the set of degrees of freedom of QED too.

\subsection{Elimination of longitudinal  component}

 This elimination is fulfilled by the choice of the "radiation variables"
 as gauge invariant functionals
 of the initial fields, \ie "dressed fields" \cite{d}
 \bea
 A^{\rm R}_\mu=A_\mu+\partial_\mu\Lambda,~~\psi^{\rm R}=e^{\imath e\Lambda}\psi,\label{d2}
 \eea
 In this case, the linear term $\partial_k A_k$ disappears
 in the Gauss law (\ref{c1})
 \be\label{1c1}
 \Delta A^{\rm R}_0=j^{\rm R}_{0}\equiv e
 \bar \psi^{\rm R}\gamma_0\psi^{\rm R}.
  \ee
 The source of the gauge-invariant {\it potential
field} $A^{\rm R}_0$
 can be only an
 electric  current $j^{\rm R}_0$; whereas
 the spatial components  of the vector field  $A^{\rm R}_k$
 coincide with the  transversal one
\be\label{kc1}
 \partial_k A^{\rm R}_k=\partial_k A^{\rm T}_k\equiv {0}.
  \ee
In this manner the frame-fixing $A_\mu=(A_0,A_k)$ combinate with
understanding of $A_0$ as a classical field and use of
  the Dirac  dressed fields (\ref{d2}) of the Gauss constraints (\ref{c1})
lead to understanding of the  variables (\ref{d2}) as
gauge-invariant functionals of the initial fields.

\subsection{Static interaction}
 Substitution of a manifest resolution
 of the Gauss constraints (\ref{c1})
  into the initial action (\ref{1e})
  calculated on constraints cause that the initial action
 can be expressed
 in terms of the gauge-invariant radiation variables
  (\ref{d2})
 \cite{d,pol}
 \bea\label{14-2}
S=
 \int d^4x \left\{\frac{1}{2}(\partial_{\mu}A^{\rm R}_k)^2
 +
 \bar \psi^{\rm R}
 [i\rlap/\partial -m]\psi^{\rm R}-
 A^{\rm R}_kj^{\rm R}_{k}+\frac{1}{2}
 j_0^{\rm R}\frac{1}{\triangle}j_0^{\rm R}\right\}
 .
 \eea
The Hamiltonian, which corresponds to this action, has the form:
 \bea\label{2-5-2}
 {\cal
 H}\!=\!\dfrac{(\Pi_k^{\rm R})^2\!+\!(\partial_jA_k^{\rm R})^2}{2}+p^{R}_{\psi}\gamma_{0}
 [i\gamma_k\partial_k\!+\!m]\psi^{\rm R}+
 A^{\rm R}_kj^{\rm R}_{k}-\frac{1}{2}
 j_0^{\rm R}\frac{1}{\triangle}j_0^{\rm R},
 \eea
  where $\Pi_{k}^{R}$, $p_\psi^{\rm R}$
  are the canonical conjugate momenta fields of the theory caluculated by standard way. By this the
  vacuum can be defined as a state with minimal energy obtained
  as the value of the Hamiltonian onto the equations of motion.
 Relativistic covariant transformations of the
 gauge-invariant fields are proved on the level of the fundamental
 operator quantization in the form of
 the Poincar\'e
 algebra generators \cite{sch2}.
 The status of the theorem of equivalence between the
 Dirac radiation variables and the Lorentz gauge formulation
 is considered in \cite{6,hpp}.

\subsection{Comparison of radiation variables with the Lorentz gauge ones}

 The static interaction and corresponding bound states
  are lost in any frame free formulation including the
  Lorentz gauge one. The action (\ref{2.3-1}) transforms into
\bea\label{2-6-1} S=\int d^4x
\left\{-\frac{1}{2}(\partial_{\mu}A^{\rm L}_\nu)^2
 +
 \bar \psi^{\rm L}
 [i\rlap/\partial -m]\psi^{\rm L}+
 A^{\rm L}_\mu j^{\rm L\mu}\right\} ,
 \eea
 where
\bea\label{2-6-2} A^{\rm L}_\mu=A_\mu+\partial_\mu\Lambda^{\rm
L},~~\psi^{\rm L}=e^{ie\Lambda^{\rm L}}\psi,~~\Lambda^{\rm
L}=-\dfrac{1}{\Box}\partial^\mu A^{\rm L}_\mu
 \eea
 are the manifest gauge-invariant functionals satisfying
 the equations of motion
 \be\label{2-6-3}
 \Box A^{\rm L}_\mu=-j^{\rm L}_\mu,
 \ee
 with the current $j^{\rm L}_\mu=-e\bar{\psi}^{\rm L}\gamma_{\mu}\psi^{\rm L}$ and
 the gauge constraints
 \be\label{2-6-4}
\partial_\mu A^{\rm L\mu}\equiv 0.
 \ee
 Really, instead of the potential (satisfying the Gauss constraints
  $\triangle A^{\rm R}_0= j^{\rm R}_0$)
 and two transverse variables
 in QED in terms of the
 radiation variables (\ref{d2}) we have here  three independent  dynamic
 variables,
 one of which $A^{\rm L}_0$ satisfies the equation
 \be\label{2-6-5}
\Box A_0^{\rm L}= -j_0,
 \ee
   and
 gives a negative contribution
 to the energy.

 We can see that there are two distinctions of the
 ``Lorentz gauge formulation'' from the
 radiation variables.
 The first is the lost of Coulomb poles (\ie static
 interactions). The second is the treatment of the time component
 $A_0$ as an independent variable with the negative contribution
 to the energy; therefore, in this case,
 the vacuum as the state with the
 minimal energy is absent.  In other words,
 one can say that the static interaction
 is the consequence of the vacuum postulate
 too. The inequivalence between the radiation variables
  and the Lorentz ones
 does not mean
violation
 of the gauge invariance,
 because both the variables  can be defined as
 the gauge-invariant functionals of the initial gauge fields
   (\ref{d2}) and (\ref{2-6-2}).

In order to demonstrate the inequivalence between the radiation
variables
  and the Lorentz ones, let us consider
 the electron-positron scattering amplitude
$T^R=\langle e^+,e^-|\hat S|e^+,e^-\rangle$. One can see that the
Feynman rules in the radiation gauge give the amplitude in terms of
the current $j_\nu=\!\bar e \gamma_\nu e$
 \bea\label{1wr}
 T^R=\frac{j^2_0}{\mathbf{q}^2}+
 \left(\delta_{ik}-\dfrac{q_iq_k}{\mathbf{q}^2}\right)
 \frac{j_ij_k}{q^2+i\varepsilon}\equiv\frac{-j^2}{q^2+i\varepsilon}+
 \fbox{$\dfrac{(q_0j_0)^2-
 (\mathbf{q}\cdot\mathbf{j})^2}{\mathbf{q}^2[q^2+i\varepsilon]}$}~.
 \eea
 This amplitude coincides with the Lorentz gauge one,
 \be
 \label{2wr}
 T^L =
 -\frac{1}{q^2+i\varepsilon}
 \left[j^2-\fbox{$\dfrac{(q_0j_0-
 \mathbf{q}\cdot\mathbf{j})^2}{q^2+i\varepsilon}$}\,\,\right]~,
 \ee
 when the box terms  in Eq. (\ref{1wr}) can be
 eliminated. Thus, the Faddeev equivalence theorem \cite{f1} is valid,
  if the currents
 are conserved
 \be \label{3wr}
 q_0{j}_0-\mathbf{q}\cdot\mathbf{j}=qj=0,
 \ee
 But for the action with the external sources   the currents
 are not conserved. Instead of the classical conservation laws
 we have the Ward--Takahashi identities for
 Green functions, where the currents are not conserved
\be \label{3wr1}
 q_0{j}_0-\mathbf{q}\cdot\mathbf{j}\not =0.
 \ee
 In particular, the
Lorentz gauge perturbation theory (where the propagator has only the
light cone singularity $q_{\mu}q^{\mu}=0$) can not describe
instantaneous Coulomb atoms; this perturbation theory contains only
the Wick--Cutkosky bound states whose spectrum is not observed in
the Nature.

Thus, we can give a response to the question: What are new physical
results that following from the
 Hamiltonian approach to QED in comparison with the
 frame free Lorentz gauge formulation? In the framework of the perturbation
 theory, the Hamiltonian presentation of  QED contains the static
 Coulomb interaction  (\ref{1wr})  forming
 instantaneous bound states observed in the nature; whereas
 all  frame free formulations lose this static interaction
 together with instantaneous bound states, in the lowest order of
 perturbation theory on retarded interactions
   called the radiation correction. Nobody proves that
   the sum of these retarded radiation correction with the
   light-cone singularity    propagators (\ref{2wr}) can restore the
   the Coulomb interaction that was removed from
   propagators (\ref{1wr})
   by the hands
   on the level of the action.

\section{The Hamiltonian approach to a massive vector theory}

\subsection{Lagrangian and reference frame}

The classical Lagrangian of massive QED is
\begin{equation}
\label{3-1-1} {\cal L}=
-\frac{1}{4}F_{\mu\nu}F^{\mu\nu}+\frac{1}{2}M^2V_\mu^2+
\bar\psi(i\rlap/\partial-m)\psi +V_{\mu}j^{\mu}~,
\end{equation}
In a fixed reference frame this Lagrangian takes the form
\begin{eqnarray}
 \label{3-1-2}\mathcal{L}\!=\!\frac{(\dot V_k\!-\!\partial_kV_0)^2\!-\!
 (\partial_jV_k^{\rm T})^2\!+\!M^2(V_0^2\!-\!V_k^2)}{2}\!+\!\bar\psi(i\rlap/\partial\!-\!m)\psi\!+\!V_{0}j_{0}\!-\!V_{k}j_{k},
\end{eqnarray}
where $\dot V=\partial_0V$ and $V_k^{\rm T}$ is the transverse
component defined by the action of the projection operator given in
Eq. (\ref{lc1}). In contrast to QED this action is not invariant
with respect to gauge transformations. Nevertheless,
 from the Hamiltonian viewpoint the massive theory
has the same problem as QED. The time component of the massive boson
has vanish canonical momentum.

\subsection{Elimination of time component}

In \cite{hpp} one supposed to eliminate the time component from the
set of degrees of freedom like the Dirac approach to QED, \ie using
the action principle. In the massive case it produce the equation of
motion
\begin{equation}\label{3-1-3}
(\triangle-M^2)V_0=\partial_i\dot{V}_i+j_0.
\end{equation}
which is understood as constraints and has solution \bea
\label{3-1-4} V_0
=\left(\frac{1}{\triangle-M^2}\,\partial_i{V}_i\right)^{\cdot}
+\frac{1}{\triangle-M^2}\,j_0.\eea In order to eliminate the time
component, let us insert (\ref{3-1-4}) into the Lagrangian
(\ref{3-1-2}) \cite{d,hpp}:
\begin{eqnarray} \nonumber
 {\cal L}\!&=&\!\frac{1}{2}\left[(\dot V_k^{\rm
 T})^2\!+\!V_k^{\rm T}(\triangle\!-\!M^2)V^{\rm T}_k
 \!+\!j_0\frac{1}{\triangle\!-\!M^2}j_0\right]\!+\!\bar\psi(i\not{\!\partial}\!-\!m)\psi\!-\!V^{\rm T}_kj_k\\\label{3-2-7}
                \!&+&\! \frac{1}{2}\left[\dot V_k^{\rm ||}
 M^2\frac{1}{\triangle\!-\!M^2}\dot V_k^{\rm ||}\!-\!M^2
 (V_k^{\rm ||})^2
 \right]\!-\!V^{\rm
 ||}_kj_k\!+\!
j_0\frac{1}{\triangle\!-\!M^2}\partial_k \dot V_k^{\rm ||},
\end{eqnarray}
where we decomposed the vector field
 $V_k=V_k^{\rm T}+V_k^{\rm ||}$ by means of the  projection
 operator by analogy to (\ref{lc1}). The last two terms are the contributions of the longitudinal
component only. This Lagrangian contains the longitudinal component
  which is  the dynamical variable described by the bilinear term.
  Now we propose the following transformation
 \bea
\bar\psi(i\not{\!\partial}\!-\!m)\psi\!-\!V^{\rm ||}_kj_k\!+\!j_0
\frac{1}{\triangle\!-\!M^2}\partial_k \dot V_k^{\rm
||}=\bar\psi^{\rm R}(i\not{\!\partial}\!-\!m)\psi^{\rm R}\!-\!V^{\rm
R ||}_kj_k\label{3-2-8b},
 \eea
where
  \bea\label{3-2-9}
  V^{\rm R||}_k&=&V^{\rm||}_k-\partial_k\frac{1}{\triangle-M^2}\,\partial_i{V}_i=
-M^2\frac{1}{\triangle-M^2}V^{\rm ||}_k~,\\
 \psi^{\rm R}&=&\exp\left\{-ie\frac{1}{\triangle-M^2}\,\partial_i{V}_i\right\}\psi
 \eea
are the radiation-type variables,
 removes the linear term $\partial_i\dot{V}_i$ in the Gauss law
 (\ref{3-1-3}). If the mass $M\not = 0$, one can
 pass from the initial variables $V^{\rm ||}_k$
 to the radiation ones $V^{\rm R ||}_k$ by the change
 \be\label{3-2-9a} V^{\rm ||}_k= \hat{Z}V^{\rm R ||}_k,~~\hat{Z}=\frac{M^{2}-\triangle}{M^{2}}
  \ee
Now the Lagrangian (\ref{3-2-7}) goes into
 \begin{eqnarray}\label{3-2-10}
 {\cal L}\!&=&\!\frac{1}{2}\left[(\dot V_k^{\rm
 T})^2\!+\!V_k^{\rm T}(\triangle\!-\!M^2)V^{\rm T}_k
 \!+\!j_0\frac{1}{\triangle\!-\!M^2}j_0\right]\!+\!\bar\psi^{\rm R}(i\not{\!\partial}\!-\!m)\psi^{\rm R}\nonumber\\
 \!&+&\!\frac{1}{2}\left[\dot V_k^{\rm R\rm ||}
 \hat{Z}\dot V_k^{\rm R\rm ||}\!+\!
 V_k^{\rm R\rm ||}(\triangle\!-\!M^{2})\hat{Z}V_k^{\rm R\rm ||}
 \right]\!-\!V^{\rm T}_kj_k\!-\!V^{\rm R ||}_kj_k.
\end{eqnarray}
The Hamiltonian corresponding to this Lagrangian can be construct in
the standard canonical way. Using rules of the Legendre
transformation and canonical conjugate momenta $\Pi_{V^{\rm T}_k}$,
$\Pi_{V^{\rm R ||}_k}$, $\Pi_{\psi^{R}}$ as a result we obtain
\begin{eqnarray}\label{3-2-14}
\cal{H}\!&=&\!\frac{1}{2}\left[\Pi_{V^{\rm T}_k}^{2}\!+\!V_k^{\rm
T}(M^2\!-\!\triangle)V^{\rm T}_k
 \!+\!j_0\frac{1}{M^2\!-\!\triangle}j_0\right]\!-\!
 \Pi_{\psi^{R}}\gamma_0(i\gamma_{k}\partial_{k}\!+\!m)\psi^{\rm R}\nonumber\\
 \!&+&\!\frac{1}{2}\left[\Pi_{V^{\rm R ||}_k}\hat{Z}^{-1}\Pi_{V^{\rm R
 ||}_k}\!+\!
 V_k^{\rm R\rm ||}(M^{2}\!-\!\triangle)\hat{Z}V_k^{\rm R\rm ||}\right]\!+\!V^{\rm T}_kj_k\!+\!V^{\rm R
 ||}_kj_k.
\end{eqnarray}
One can be convinced \cite{hpp} that the corresponding
 quantum system has a vacuum as a state with
 minimal energy and correct relativistic transformation
 properties.

\subsection{Quantization}

We start the quantization procedure from the canonical quantization
by using the following equal time canonical commutation relations
(ETCCRs):
\begin{eqnarray}\label{3-4-1}
 \left[\hat{\Pi}_{V^{\rm T}_k},\hat{V}^{\rm
 T}_k\right]=i\delta_{ij}^{\rm T}\delta^3(\mathbf{x}-\mathbf{y}),\\
\left[\hat{{\Pi}}_{V^{\rm R ||}_k}, \hat{V}^{\rm R ||}_k\right]=
 i\delta_{ij}^{||}\delta^3(\mathbf{x}-\mathbf{y}).
\end{eqnarray}
The Fock space of the theory is building by the ETCCRs
\begin{eqnarray}
\left[{a^{-}_{(\lambda)}\left({\pm
k}\right),a_{(\lambda')}^{+}\left({\pm k'}\right)}\right]&=&\delta
^{3}\left({{\bf k}-{\bf k'}}\right)\delta_{(\lambda)(\lambda')};\\
\left\{b^{-}_\alpha\left({\pm k}\right),b_{\alpha'}^{+}\left({\pm
k'}\right)\right\}&=&\delta^{3}\left({{\bf k}-{\bf
k'}}\right)\delta_{\alpha\alpha'};\\
\left\{{c^{-}_\alpha\left({\pm k}\right),c_{\alpha'}^{+}\left({\pm
k'}\right)}\right\}&=&\delta^{3}\left({{\bf k}-{\bf
k'}}\right)\delta_{\alpha\alpha'}.
\end{eqnarray}
with the vacuum state $|0\rangle$ defined by the relations:
 \be \label{3-4-7}
 a_{(\lambda)}^-|0\rangle=b_\alpha^-|0\rangle=c_\alpha^-|0\rangle=0.
 \ee
The field operators has the Fourier decompositions in the plane
waves basis
\begin{eqnarray}\label{1.8}
 V_j\left(x\right)&=&
 \int[d\mathbf{k}]_{v}~
 \epsilon_{j}^{(\lambda)}{\left[{a_{(\lambda)}^{+}\left({\omega,{\bf k}}
 \right)e^{-i\omega t + i{\bf kx}}\!+\!a_{(\lambda)}^{-}
 \left({\omega,-{\bf k}}\right)e^{i\omega t -i{\bf kx}}}\right]},\\
 \psi\left(x\right)&=&\sqrt{2m_{s}}\!\int[d\mathbf{k}]_{s}{\left[{b^+_\alpha\left(k\right)u_\alpha
 e^{-i\omega t + i{\bf kx}}+c^-_\alpha\left({-k}\right)\nu _\alpha e^{i\omega t -i{\bf kx}}}\right]},\\
 \psi^{+}\left(x\right)&=&\sqrt{2m_{s}}\!\int[d\mathbf{k}]_{s}{\left[{b^-_\alpha\left(k\right)u_\alpha^{+}e^{i\omega t -i{\bf kx}}+c^+_\alpha
 \left({-k}\right)\nu _\alpha^{+}e^{-i\omega t +i{\bf kx}}}\right]},
\end{eqnarray}
with the integral measure
$[d\mathbf{k}]_{v,s}=\dfrac{1}{\left({2\pi}\right)^{3/2}}\dfrac{{d^3\bf
k}}{{\sqrt{2\omega_{v,s}(\bf k)}}}$ and the frequency of
oscillations
$\omega_{v,s}(\mathbf{k})=\sqrt{\,\mathbf{k}^2+m^2_{v,s}}$. One can
define the vacuum expectation values of the
 instantaneous products of the field operators
\bea \label{3-4-8}
 V_i(t,\vec x)V_j(t,\vec y)&=&~:V_i(t,\vec x)V_j(t,\vec y):
 +~\langle V_i(t,\vec x)V_j(t,\vec y) \rangle,\\
 \overline{\psi}_\alpha(t,\vec x) \psi_\beta(t,\vec y)&=&~:
 \overline{\psi}_\alpha(t,\vec x) \psi_\beta(t,\vec y):
 +~\langle \overline{\psi}_\alpha(t,\vec x) \psi_\beta(t,\vec y),
 \eea
where
 \bea \label{3-4-9}
 \langle V_i(t,\vec x)V_j(t,\vec y) \rangle=\frac{1}
 {(2\pi)^3}\int\frac{d^3\bf{k}}{2\omega_v(\bf{k})}
 \sum\limits_{(\lambda)}^{}
 \epsilon_{i}^{(\lambda)}\epsilon_{j}^{(\lambda)}e^{-i\bf{k}(\bf{x}-\bf{y})},
\\\label{3-4-9a}
 \langle \overline{\psi}_\alpha(t,\vec x) \psi_\beta(t,\vec y) \rangle=
 \frac{1}{(2\pi)^3}\int\frac{d^3\bf{k}}{2\omega_s(\bf{k})}
 (\mathbf{k}\vec{\gamma}+m)_{\alpha\beta}\,e^{-i\bf{k}(\bf{x}-\bf{y})}
\eea
 are the  Pauli -- Jordan functions.

\subsection{Propagators and condensates}

The vector field in the Lagrangian (\ref{3-2-10}) is given by the
formula
\begin{equation}\label{3-5-1}
V^{\rm R}_i= \left[\delta_{ij}^{\rm T}+\hat Z^{-1}\delta_{ij}^{\rm
||}\right]V_j=V_i^{\rm T}+\hat Z^{-1}V_i^{\rm ||}.
\end{equation}
Use of this give us the propagator of the massive vector field in
radiative variables
\begin{eqnarray}\label{3-5-2}
D^R_{ij}(x\!-\!y)\!=\!\langle
0|TV^R_{i}(x)V^R_{j}(y)|0\rangle\!=\!\!-\!i\int
\frac{d^4q}{(2\pi)^4}\frac{e^{\!-\!iq\cdot
(x\!-\!y)}}{q^2\!-\!M^2\!+\!i\epsilon}\!\left(\delta_{ij}\!-\!\frac{q_{i}q_{j}}{\mathbf{q}^2\!+\!M^2}\right)~.
\end{eqnarray}
 Together with the instantaneous  interaction described by
  the current--curent term in
 the Lagrangian (\ref{3-2-10})  this propagator
 leads to the amplitude
\begin{equation}\label{3-2-11}
T^{\rm R}= D^{\rm R}_{\mu\nu}(q)\widetilde{j}^\mu \widetilde{j}^\nu
=
\frac{\widetilde{j}_0^2}{\mathbf{q}^2+M^2}+\left(\delta_{ij}-\frac{q_i
q_j}{\mathbf{q}^2+M^2}\right) \frac{\widetilde{j}_i
\widetilde{j}_j}{{q}^2-M^2+i\epsilon}~
\end{equation}
 of the current-current interaction, which differs from the acceptable one
\begin{equation}\label{3-1-8}
 T^{\rm L}=\widetilde{j}^{\mu}D^{\rm L}_{\mu\nu}(q)\widetilde{j}^{\nu}=
-\widetilde{j}^{\mu}\frac{g_{\mu\nu}-\dfrac{q_\mu q_\nu}{M^2}
}{q^2-M^2+i\epsilon}\widetilde{j}^{\nu}.
\end{equation}
The amplitude given by Eq. (\ref{3-2-11})  is the generalization of
the  radiation amplitude  in QED. As it was shown in  \cite{hpp},
the Lorentz transformations of classical radiation variables
coincide with the  quantum ones  and they both (quantum and
classical) correspond to the transition to another Lorentz frame of
reference distinguished by another time-axis, where the relativistic
covariant propagator takes the form
\begin{eqnarray}\label{3-5-4}
D^{R}_{\mu\nu}(q|n)\!=\!
\!-\!\frac{1}{q^2\!-\!M^2\!+\!i\epsilon}\left[g_{\mu\nu}\!-\!
\dfrac{n_{\mu}n_{\nu}(qn)^2\!-\![q_{\mu}\!-\!n_{\mu}(qn)][q_{\nu}\!-\!n_{\nu}(qn)]}
 {M^2\!+\!|q_{\mu}\!-\!n_{\mu}(qn)|^2}\right]\label{3-5-4a}
\end{eqnarray}
where $n_{\mu}$ is determined by the external states. Remember that
 the conventional local field massive vector propagator
 takes the form  (\ref{3-1-8})
 \begin{equation}\label{3-5-5}
 D^L_{\mu\nu}(q)=-
\dfrac{g_{\mu\nu}-\dfrac{q_\mu q_\nu}{M^2}}{q^2-M^2+i\epsilon}~.
\end{equation}
In contrast to this conventional  massive vector propagator the
 radiation-type propagator (\ref{3-5-4})  is regular in the
limit $M\rightarrow 0$ and is well behaved for large momenta,
whereas the propagator (\ref{3-5-5}) is singular. The radiation
amplitude (\ref{3-2-11}) can be rewritten   in the alternative form
\begin{equation}
T^{\rm R}=-\frac{1}{q^2-M^2+i\epsilon}\left[\widetilde{j}_{\nu}^2
+\frac{(\widetilde{j}_iq_i)^2-(\widetilde{j}_0q_0)^2}{\vec{q}^2+M^2}\right]~,
\label{mvecprop2}
\end{equation}
for comparison with the conventional amplitude defined by the
propagator (\ref{3-5-5}). One can find that for a massive vector
field coupled to a conserved current
$(q_{\mu}\widetilde{j}^{\mu}=0)$ the collective current-current
interactions mediated by the radiation propagator  (\ref{3-5-4}) and
by the conventional propagator (\ref{3-5-5}) coincide
\begin{equation}\label{3-5-52}
\widetilde{j}^{\mu}D^{\rm R}_{\mu\nu}\widetilde{j}^{\nu}=
\widetilde{j}^{\mu}D^{\rm L}_{\mu\nu}\widetilde{j}^{\nu}=T^{\rm L}~.
\end{equation}
 If the  current is not conserved $\widetilde{j}_0q_0\not =\widetilde{j}_kq_k$,
 the collective
 radiation field variables
 with the propagator (\ref{3-5-4})
 are inequivalent to   the initial local  variables
 with the propagator  (\ref{3-5-5}),  and the amplitude
 (\ref{3-2-11}). The amplitude (\ref{3-5-52}) in the Feynman gauge is
 \be\label{3-5-52a}
 T^{\rm L} =
 -\frac{j^2}{q^2-M^2+i\varepsilon},
 \ee
and corresponds to the Lagrangian
 \be\label{brst}
\mathcal{L}_{F}=\frac{1}{2}(\partial_{\mu}V_{\mu})^{2}-j_{\mu}V_{\mu}+
\frac{1}{2}M^2 V_\mu^2 \ee
 In this theory the time component has a negative contribution
 to the energy. By this correct defined
 vacuum state could not exist. Nevertheless, the vacuum expectation
 value $\langle V_\mu(x)V_\mu(x) \rangle$ coincides with the values
 for two  propagators (\ref{3-5-4a}) and
 (\ref{3-5-5})
 because in both these propagators the longitudinal part 
 do not contribute, if one treats they as derivatives of constant
 like
 $\langle \partial V_\mu(x)V_\mu(x) \rangle
 =\partial \langle  V_\mu(x)V_\mu(x) \rangle=0$.
 In this case we have
\bea \label{v3-4-8}
 \langle V_\mu(x)V_\mu(x) \rangle &=&-\frac{2}
 {(2\pi)^3}\int\frac{d^3\bf{k}}{\omega_v(\bf{k})}\simeq 2 L_v^2(M_v),
 \\\label{s3-4-8}
 \langle \overline{\psi}_\alpha(x)  \psi_\alpha(x)\rangle&=&-
 \frac{m_s}{(2\pi)^3}\int\frac{d^3\bf{k}}{\omega_s(\bf{k})}=m_s
 L_s^2(m_s),
 \eea
where $m_s$, $M_{v}$ are masses of the spinor and vector fields and
$L^2_{s,v}$ are values of the integrals that coincide in the
massless limit \mbox{$L_v^2=L_s^2\sim\int d^3k /|\vec k|$}.

\section{On the Hamiltonian presentation of SM}
\subsection{The SM action}
 The elementary particle physics which is successfully
described in frames of the Standard Model. As an example, we are
going to study the electroweak sector of SM, but analogously
procedure can be formulate with strong interactions presence. The
electroweak sector of SM is described by the Yang--Mills theory
\cite{ym} constructed on the symmetry group ${SU(2)}\times{U(1)}$.
It is known as the Glashow-Weinberg-Salam theory of electroweak
interactions \cite{weak}. We include the Higgs boson existence. The
action of the SM in the electroweak sector, with presence of the
 Higgs field can be write in the form
 \be\label{1-sm}
 S_{\rm SM}=\int d^4x
 {\cal L}_{\rm SM}=\int d^4x\sqrt{-g}\left[{\cal L}_{\rm Inv}
 +{\cal L}_{\rm Higgs}
 \right],
 \ee
where
 \bea\label{M}
{\cal L}_{\rm Inv}&=&-\frac{1}{4}G^a_{\mu\nu}
G^{\mu\nu}_{a}-\frac{1}{4}F_{\mu\nu} F^{\mu\nu}+\\
\nonumber&+&\sum_s\bar{s}_1^{R}\imath
\gamma^{\mu}\left(D_{\mu}^{(-)}
 +\imath
g^{\prime}B_{\mu}\right)s_1^{R}+\sum_s\bar{L}_s\imath
\gamma^{\mu}D^{(+)}_{\mu}L_s,\\\label{Ma}
{\mathcal{L}_{\rm{Higgs}}}&=&
{\partial_\mu\phi\partial^\mu\phi} -\phi\sum_sf_s\bar s
s+\frac{{\phi^2}}{4}\sum_{\rm v} g^2_{\rm v}V^2-
\underbrace{{\lambda}\left[\phi^2-\Phi_0^2\right]^2}_{V_{\rm Higgs}}
\eea
 are the Higgs field independent and the Higgs field dependent
 parts of the Lagrangian respectively;
 here $G^a_{\mu\nu}=\partial_{\mu}A^{a}_{\nu}-\partial_{\nu}A^{a}_{\mu}+g\varepsilon_{abc}A^{b}_{\mu}A^{c}_{\nu}$
is the field strength of non-Abelian $SU(2)$ fields (that give weak
interactions) and
\mbox{$F_{\mu\nu}=\partial_{\mu}B_{\nu}-\partial_{\nu} B_{\mu}$} is
the field strength of Abelian $U(1)$ (electromagnetic interaction)
ones,
$D_{\mu}^{(\pm)}=\partial_{\mu}-i{g}\frac{\tau_{a}}{2}A^a_{\mu}\pm\frac{i}{2}g^{\prime}B_{\mu}$
are the covariant derivatives, $\bar L_s=(\bar s_1^{L}\bar s_2^{L})$
are the lepton doublets, $g$ and $g'$ are the Weinberg coupling
constants. The quantities coupled with the Higgs field are equal
\begin{eqnarray}\label{1-6a}
 \sum_s f_s\bar s s&\equiv&\sum_{s=s_1,s_2} f_{s}\left[\bar s_{sR}s_{sL}+\bar s_{sL}s_{sR}\right],\\\label{w1-6a}
\frac{1}{4}\sum_{\rm v=W_1,W_2,Z} g^2_{\rm
v}V^2&\equiv&\frac{g^{2}}{4}W_\mu^{+}W^{-\mu}+\frac{g^{2}+g'^{2}}{4}Z_{\mu}Z^{\mu}
\end{eqnarray}
 are the mass-like terms  of  fermions  and W-,Z-bosons,
 respectively. Measurable gauge bosons $W^+_{\mu},~W^-_{\mu},~Z_{\mu}$ are defined by the relations:
\bea
W_{\mu}^{\pm}&\equiv&{A}_{\mu}^1\pm{A}_{\mu}^2,\\
Z_{\mu}&\equiv&-B_{\mu}\sin\theta_{W}+A_{\mu}^3\cos\theta_{W},\\
\tan\theta_{W}&=&\frac{g'}{g},\eea where $\theta_{W}$ is the
Weinberg angle. The crucial meaning has a distribution of the Higgs
field $\Phi$ on the zero Fourier harmonic
 \be\label{h0-1}
 \langle\phi\rangle=\frac{1}{V_0}\int d^3x \phi
 \ee and the non zero ones $h$, which we will call name
 the Higgs boson:
  \be\label{h-1}
 \phi=\langle\phi\rangle+\frac{h}{\sqrt{2}},~~~~\int d^3x h=0.
 \ee
In the acceptable  way, $\langle\phi\rangle$ satisfies the
 particle vacuum classical equation ($h=0$)
 \be\label{hc-1}
 \frac{\delta V_{\rm Higgs}(\langle\phi\rangle)}
 {\delta\langle\phi\rangle}=4{\langle\phi\rangle}
 [{\langle\phi\rangle}^2-\Phi^2_0]=0
 \ee
 that has two solutions
\be\label{hc-2}
 {\langle\phi\rangle}_1=0,~~~~~~~~~~~~~~~
 {\langle\phi\rangle}_2=\Phi_0\not =0.
 \ee
 The second solution corresponds to the spontaneous vacuum symmetry
 breaking
 that determines the masses of all elementary particles
 \bea
\label{1W-1}M_{W}&=&\frac{\Phi_0}{\sqrt{2}}g\\\label{1Z-1}
M_{Z}&=&\frac{\Phi_0}{\sqrt{2}}\sqrt{g^2+g'^2}
\\\label{1s-1}m_{s}&=& \Phi_0 y_s
\eea according to  definitions of the masses of vector (v) and
fermion (s) particles
 \be\label{vs-1}
{\cal L}_{\rm mass~ terms}= \frac{M_v^2}{2}V_\mu V^\mu-m_s\bar s s.
 \ee

\subsection{Hamiltonian approach to SM}

 The  accepted SM (\ref{1-sm})
 is bilinear with
 respect to the time components of the vector fields $V^K_0=(A_0,Z_0,W_0^{+},W_0^{-})$
 in the ``comoving frame'' $n^{\rm cf}_\mu=(1,0,0,0)$
 \be\label{sm4}
 S_{V}=\int d^4x \left[\frac{1}{2}V^K_0\hat L^{KI}_{00}V^I_0+V^K_0J^K+...
 \right]~,
 \ee
  where $\hat L^{KI}_{00}$ is
 the matrix of differential operators. Therefore, the Dirac approach
to SM can be realized. This means that the
 problems of the reduction and
  diagonalization of the set of the Gauss laws are solvable, and
  the Poincar\'e algebra of gauge-invariant observables can be proved \cite{hpp}.
In any case,  SM in the lowest order of  perturbation theory is
reduced to
 the sum of the Abelian massive vector fields, where
  Dirac's  radiation variables was considered
 in Sections 3.
\subsection{The dynamic Higgs effect}

 The Hamiltonian approach to SM leads
 fundamental operator quantization that allows
a possibility of the dynamic spontaneous symmetry breaking
 based on  in
the Higgs potential  (\ref{Ma}), where instead of a dimensional
parameter $\Phi_0$ we substitute the zero Fourier harmonic
(\ref{h0-1}) \bea\label{0-Ma}
{\mathcal{L}_{\rm{Higgs}}}&=&
{\partial_\mu\phi\partial^\mu\phi} -\phi\sum_sf_s\bar s
s+\frac{{\phi^2}}{4}\sum_{\rm v} g^2_{\rm v}V^2-
\underbrace{{\lambda}\left[\phi^2-\langle\phi
\rangle^2\right]^2}_{V_{\rm Higgs}}. \eea

 The vacuum Lagrangian for the zero Fourier harmonic takes the form
 \bea\label{0-Ma-1}
{\mathcal{L}^{\lambda}_{\rm{Higgs}}}&=&
(\partial_0\langle\phi\rangle)^2 -V^{\rm eff}_{\rm
Higgs},\\\label{01-Ma-1} V^{\rm
eff}_{\rm{Higgs}}&=&\langle\phi\rangle\sum_sf_s<\!\!\bar s
s\!\!>+\frac{{\langle\phi\rangle^2}}{4}\sum_{\rm v} g^2_{\rm
v}<\!\!V^2\!\!>- 2{\lambda}{\langle\phi\rangle^2}<\!\!h^2\!\!>, \eea
 where $<V^2>,<\bar ss>$ are condensates given by Eqs. (\ref{v3-4-8}),
  (\ref{s3-4-8}), and
 \bea\label{h3-4-8}\langle h^2\rangle =\langle h(x)h(x) \rangle
=\frac{1}
 {2(2\pi)^3}\int\frac{d^3\bf{k}}{\sqrt{m^2_{h}+\bf{k}^2}}= \frac{1}{2}
 L_h^2
 \eea
 is the Higgs condensate and
 \bea
\label{0W-1}M^2_{W}&=&\frac{\langle\phi\rangle^2}{{2}}g^2,\\\label{0Z-1}
M^2_{Z}&=&\frac{\langle\phi\rangle^2}{{2}}({g^2+g'^2}),
\\\label{0s-1}m_{s}&=& \langle\phi\rangle\, y_s,\\
\label{0h-1}
 m^2_h&=&4\lambda\langle\,\phi\rangle^2
\eea are  definitions of the masses of vector (v), fermion (s) and
 Higgs (h) particles.
 The zero Higgs mode
$\langle\phi\rangle$ satisfies the vacuum classical equation
 \be\label{0hc-1}
 \partial^2_0\langle\phi\rangle+\frac{\delta V^{\rm eff}_{\rm Higgs}(\langle\phi\rangle)}
 {\delta\langle\phi\rangle}=0
 \ee
 that takes a form
\bea\label{0-Ma-2}
\partial^2_0\langle\phi\rangle+\sum_sf_s<\!\!\bar s
s\!\!>-\frac{{\langle\phi\rangle}}{2}\sum_{\rm v} g^2_{\rm
v}<\!\!V^2\!\!>+
\underbrace{4{\lambda}{\langle\phi\rangle^2}<\!\!h^2\!\!>}_{V_{\rm
Higgs}}=0. \eea
 Finally using the definitions of the condensates and
 masses (\ref{0W-1}), (\ref{0Z-1}),(\ref{0s-1}),(\ref{0h-1}) we
 got the equation of motion
\bea\label{0-Ma-3}
\langle\phi\rangle\partial^2_0\langle\phi\rangle=\sum_sm_s^2
L_s^2-2\sum_{\rm v} M_{\rm v}^2L^2_{\rm v}-\frac{1}{2}m_h^2L^2_{\rm
h} . \eea
 In the class of the constant solutions $\partial^2_0\langle\phi\rangle\equiv 0$
 this equation has two solutions
\be\label{hc-2}
 {\langle\phi\rangle}_1=0,~~~~~~~~~~~~~~~
 {\langle\phi\rangle}_2=\Phi_0\not =0.
 \ee
\bea\label{31-t}
 m^2_h=2\sum_{s=s_1,s_2}\frac{L_s^2}{L_h^2}m^2_s
 -4\frac{[2M_W^2L_W^2+M_Z^2L_Z^2]}{L_h^2}.
 \eea
In the minimal SM \cite{db}, a three color t-quark dominates $\sum_s
m_s^2\simeq 3m_t^2$ because
  contributions of other fermions $\sum_{s\not=t} m_s^2/2m_t\sim 0.17$ GeV
 are very small. The large cut-off limit $L_W^2=L_Z^2=L_f^2=L_h^2=L^2$
 leads to  the
 value of the Higgs mass \bea\label{41-t}
 m_h=\sqrt{6m^2_t-4[2M_W^2+M_Z^2]}=311.6\pm 8.9 \mbox{\rm GeV},
 \eea
   if we substitute  the PDG data on the  values of masses of bosons
 $M_W=80.403\pm0.029$ GeV, $M_Z=91.1876\pm0.00021$ GeV,
 and t-quark $m_t=174.2\pm 3.3$ GeV.

\subsection{The static interaction mechanism of
  the enhancement of the $\triangle T=1/2$ transitions}

\begin{figure}
  \begin{center}
  \includegraphics[scale=0.6]{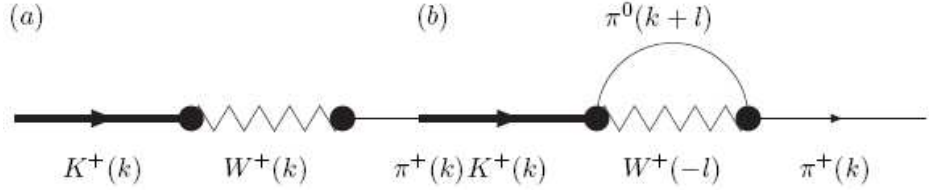}
  \caption{Axial (a) and vector (b) current contribution into $K^+\to
\pi^+$ transition} \label{1ac}
  \end{center}
\end{figure}

 Let us consider the $K^+\to \pi^+$ transition amplitude
\be\nonumber
 \left\langle\pi^+\left|-i\int d^4xd^4y
 J^\mu(x)D_{\mu\nu}^W(x-y)J^\nu(y)\right|K^+\right\rangle
 =i(2\pi)^{4}\delta^{4}(k-p)G_{\rm EW}\Sigma(k^{2})
 \ee
 in the first order of the EW perturbation theory in the Fermi
 coupling constant \be\label{g}
  G_{\rm EW}= \frac{\sin\theta_{C} \cos\theta_{C}}{8
  M^{2}_{W}}\frac{e^{2}}{\sin^{2}\theta_{W}}\equiv
  \sin\theta_{C} \cos\theta_{C}\frac{G_F}{\sqrt{2}},
 \ee
 comparing two different W-boson field propagators,
 the accepted Lorentz (L)
 propagator (\ref{3-5-5})
 and
 the  radiation (R) propagator (\ref{3-5-4a}). 
 These propagators give the expressions
 corresponding to the diagrams in Fig. \ref{1ac}
\bea\label{kp2}
  \Sigma^R(k^{2})&=&\!2F^2_{\pi}{k^2}
  -2i\int \!\! \frac{d^4q{M_W^2}}{(2\pi)^4}
 \frac{k^2+(k_0+q_0)^2}{(|\vec q|^2+M_W^2)[(k+q)^2
 \!-\!m^2_\pi+i\epsilon]}, \\
 \label{fkp2}
  \Sigma^L(k^{2})&=&2F^2_{\pi}{k^2}
 +2i\int \frac{d^4q{M_W^2}}{(2\pi)^4}
 \frac{(2k_\mu+q_\mu)D^{L}_{\mu\nu}(-q)(2k_\nu+q_\nu)}{(k+q)^2
 -m^2_\pi+i\epsilon}.
 \eea
 The versions R and L coincide in the case of the axial contribution
 corresponding to the first diagram in Fig. \ref{1ac},
 and they both reduce to the static interaction contribution
 because
$$
k^\mu k^\nu D^F_{\mu\nu}(k)\equiv k^\mu k^\nu
D^R_{\mu\nu}(k)=\frac{k_0^2}{M^2_W}.
$$

%

 However, in the case of the vector contribution
 corresponding to the second diagram in Fig.  \ref{1ac}
 the radiation version differs from the  Lorentz gauge version
  (\ref{3-5-5})\footnote{The Faddeev equivalence theorem \cite{f1}
 is not valid, because
  the vector current $J_\mu = K\partial_\mu \pi-\pi\partial_\mu K$
  becomes the vertex $\Gamma_\mu = K\partial_\mu D_\pi-D_\pi\partial_\mu K$,
   where one of fields
  is replaced by its propagator $\Box D_\pi =\delta (x)$, and
  $\partial_\mu \Gamma^\mu \not = 0$.}.

  In contrast to the Lorentz gauge version  (\ref{3-5-5}),
  two radiation variable diagrams  in Fig. \ref{1ac} in  the
  rest kaon frame $k_\mu=(k_0,0,0,0)$
  are reduced to
  the static interaction contribution
 \be\label{kp1}
 i(2\pi)^{4}\delta^{4}(k-p)G_{\rm EW}\Sigma^R(k^{2})=
 \left\langle\pi^+\left|-i\int d^4x
 J_0(x)\frac{1}{\triangle-M_W^2}J_0(x)\right|K^+\right\rangle\nonumber
 \ee
  with the normal ordering of the pion fields which are
  at their mass-shell\footnote{The second  integral in (\ref{kp2})
   with the term  $(k_0+q_0)^2$
   really does not depend on $k^2$, and it can be removed by the mass rotation.},
   so that
 \be\label{0Rkp1}
 \Sigma^R(k^{2}) = 2k^2{F^2_{\pi}}\left[1+
  \frac{M_{W}^{2}}{F^2_{\pi}(2\pi)^3}\int\frac{d^3l}{2E_\pi(\vec{l})}
  \frac{1}{M^2_W+\vec{l}^2}\right]
 \equiv 2k^2{F^2_{\pi}}g_8.
 \ee
Here $E_{\pi}(\vec{l})=\sqrt{m^{2}_{\pi}+\vec{l}^{2}}$ is the
 energy of $\pi$-meson and $g_8$ is
 the parameter of   the enhancement of the probability
 of the axial $K^+ \to \pi^+$ transition.
 The pion mass-shell justifies
 the application of the low-energy ChPT~\cite{fpp,vp1},
 where the summation of the chiral series can be
 considered here as
  the meson form factors
  \cite{bvp,ecker,belkov01}
  $\int\dfrac{d^3l}{2E_\pi(\vec{l})}\to$
  $\int\dfrac{d^3l f^V_K(-(\vec l)^2)f^V_\pi(-(\vec l)^2)}{2E_\pi(\vec{l})}$.

 Using
  the covariant perturbation theory \cite{pv}\ developed as the
 series $\underline{J}_\mu^k(\gamma
 \oplus\xi)=\underline{J}_\mu^k(\xi)+ F^2_\pi
 \partial_\mu \gamma^k +\gamma^if_{ijk}\underline{J}_\mu^j(\xi)+O(\gamma^2)$
 with respect to quantum fields $\gamma$ added to $\xi$
 as  the product $e^{i\gamma}e^{i\xi}\equiv e^{i(\gamma \oplus\xi)}$, one can see that the normal
  ordering $$<0|\gamma^i(x)\gamma^{i'}(y)|0>=\delta^{ii'}N(\vec
  z),~~~~
  N(\vec z)=\int \frac{d^3l e^{i\vec l\cdot (\vec z)}}{(2\pi)^32E_\pi(\vec l)},$$
  where $\vec z=\vec x-\vec y$,
  in the product of the currents $\underline{J}_\mu^k(\gamma
 \oplus\xi)$
   leads to an 
effective Lagrangian with the rule $\triangle T=1/2$ 
$$
 M_W^2\int d^3z g_8(z)\frac{e^{-M_W|\vec z|}}{4\pi|\vec z|}
 [\underline{J}_\mu^j(x)\underline{J}_\mu^{j'}(z+x)
(f_{ij1}+i f_{ij2})(f_{i'j'4}-if_{i'j'5})\delta^{ii'}+h.c] ,
 $$
 where
 $g_8(|z|)= [1+\sum\limits_{I\geq 1}c^I N^I(\vec z)]
 $
 is series
 over the multipaticle intermediate states (this sum is known as
 the Volkov superpropagator \cite{vp1,S}).
 In the  limit $M_W\to \infty$,
  in the lowest order with respect to $M_W$,  the
  dependence of $g_8(|\vec z|)$ and the currents on $\vec z$
 disappears in the integral of the type of
 $$M^2_W \int d^3z \dfrac{g_8(|\vec z|)e^{-M_W|\vec z|}}{4\pi|\vec z|}=
 \int_0^{\infty} drr e^{-r}g_8({r}/{M_W})\simeq g_8(0).$$
 In the next order, the amplitudes $K^0(\bar K^0)\to \pi^0$ arise.
  Finally, we get
  the effective  Lagrangians  \cite{kp}
 \bea \label{ef1}\mathcal{L}_{(\Delta T=\frac{1}{2})}&=&
\frac{G_{F}}{\sqrt{2}}g_{8}(0)\cos\theta_{C}\sin\theta_{C}\times\\\nonumber
&&\left[(\underline{J}^1_{\mu}+i\underline{J}^2_{\mu})
(\underline{J}^4_{\mu}-i\underline{J}^5_{\mu})-
\left(\underline{J}^3_{\mu}+\frac{1}{\sqrt{3}}\underline{J}^8_{\mu}\right)
(\underline{J}^6_{\mu}-i\underline{J}^7_{\mu})+h.c.\right], \eea
\be\label{ef2} \mathcal{L}_{(\Delta T=\frac{3}{2})}=
\frac{G_{F}}{\sqrt{2}}\cos\theta_{C}\sin\theta_{C}
\left[\left(\underline{J}^3_{\mu}+\frac{1}{\sqrt{3}}\underline{J}^8_{\mu}\right)
(\underline{J}^6_{\mu}-i\underline{J}^7_{\mu})+h.c.\right]. \ee

 This result shows that the enhancement can be explained
  by static vector interaction that
  increases the $K^+\to \pi^+$ transition
 by a factor of $g_8=g_8(0)$, and yields a new term describing the
  $K^0\to \pi^0$ transition proportional to $g_8-1$.

 This Lagrangian with the fit parameter $g_8=5$
 (i.e. $g_8\sin\theta_{C} \cos\theta_{C}\simeq 1$) 
 describes the nonleptonic decays in
 satisfactory agreement with experimental data 
 \cite{vp1,kp,06}.

 Thus, for normal ordering of
 the weak static interaction in the Hamiltonian SM
 can explain the rule $\triangle T=1/2$ and universal factor $g_8$.


 On the other hand, contact character of
 weak static interaction in the Hamiltonian SM
  excludes  all retarded diagram contributions in the effective
   Chiral Perturbation
  Theory considered in \cite{da98} that destruct
  the form factor structure of the kaon radiative decay rates
 with the amplitude \be
  T_{(K^+\to\pi^+l^+l^-)}=g_8 t(q^2)
   2F^2_\pi  \sin\theta_{C} \cos\theta_{C}\dfrac{G_F}{\sqrt{2}}
 \dfrac{(k_{\mu }+p_{\mu })}{q^2}\bar{l}\gamma_{\mu}l\,\,\,
 \ee
where $q^2=(k-p)^2$, and
 \be
 t(q^2)=
 \dfrac{f^{A}_{K}(q^{2})\!+\!f^{A}_{\pi}(q^{2})}{2}
 - f^{V}_{\pi}(q^{2})+
 \Big[\!f^{V}_{K}(q^{2})\!-\!
 f^{V}_{\pi}(q^{2})\!\Big]\dfrac{m_\pi^2}{M_K^2-m_\pi^2},
 \ee
 and $f_{K}^{V}\simeq f_{\pi}^{V}(q^2)=1+M^{-2}_\rho q^2+...$,
$f_{K}^{A}(q^2)\simeq f_{\pi}^{A}(q^2)=1+M^{-2}_a
 q^2+\ldots$ are form factors determined by
 the masses of the nearest resonances for meson -- gamma --
 meson vertex.

 Therefore, the static interaction mechanism of
  the enhancement of the $\triangle T=1/2$ transitions
 predicts \cite{06}  that  the meson form factor
 resonance parameters explains the experimental values of rates of
  the radiation kaon decays  $K^+ \to
 \pi^+e^{+}e^{-}(\mu^{+}\mu^{-})$.
 Actually,  substituting of the PDG data on the resonance masses
 $M_\rho=775.8$ {\rm MeV}, $1^+(1^{--})=I^G(J^{PC})$
 and meson -- gamma -- W-boson one $M_a=984.7$ {\rm MeV},
  $1^-(0^{++})=I^G(J^{PC})$ into the decay amplitudes
 one can obtain
 decay branching fractions \cite{06}:
\begin{eqnarray}\nonumber
{\rm Br}(K^+\to\pi^+e^+e^-) &=& 2.93 \times 10^{-7},~~[2.88\pm0.13
\times 10^{-7}]_{\rm PDG }\\{\rm Br}(K^+\to\pi^+\mu^+\mu^-) &=& 0.73
\times 10^{-7},~~[0.81\pm0.14 \times 10^{-7}]_{\rm PDG }
\end{eqnarray}
 in satisfactory agreement with the experimental data \cite{ap99,pdg}.
  Thus, the
  off-mass-shell kaon-pion transition in the radiation weak kaon decays
     can be a good probe of the weak static interaction
     revealed by the
     radiation propagator (\ref{3-5-4a}) of the Hamiltonian presentation of SM.

\section{Conclusions}

\emph{Physical consequences of the Hamiltonian
   approach to the Standard Model are the weak  static interactions,
   like
    the Coulomb static interaction is
consequence of the Hamiltonian  approach in QED.} The static
interactions can be  omitted, if we
 restrict ourselves 
 by the scattering processes of the elementary particles
 where static interactions
 are not important. However, the static poles
 play crucial role in the mass-shell
 phenomena of the    bound state type,  spontaneous symmetry breaking,
  kaon - pion
 transition in the weak decays, etc.
  \emph{Static interactions follow from the  spectrality principle that means
 existence of a vacuum
 defined as a state with the minimal energy.}
  We discussed physical effects testifying  about the static
 interactions omitted by the  accepted version of SM.

 One of these effects is revealed by the loop
  meson diagrams in the low energy weak static interaction.
 These  diagrams lead to the enhancement coefficient $g_8$ in weak kaon
 decays and  the rule $\triangle T=\frac{1}{2}$.
    The loop pion diagrams
     in the Chiral Perturbation Theory  \cite{vp1} in
     the framework of the Hamiltonian approach
      with
  the weak static interaction
 lead to definite relation
  of the vector form factor with the differential radiation kaon decay
   rates in agreement with the present day PDG data \cite{06},
 in the contrast to the acceptable renormalization group
     analysis based on the Lorentz gauge formulation omitting weak static
     interaction \cite{da98}, where loop pion diagrams destroy
     the above mentioned  relation
  of the vector form factor with the differential radiation kaon decay
   rates.
 \emph{Therefore,
   the radiation kaon decays can be a good probe
   of the  weak static  interaction.}

The Hamiltonian  approach to the Standard Model and its operator
fundamental
   formulation lead to a dynamic version
of spontaneous symmetry breaking and the Schwinger -- Dyson type
equation
 provoking the sum-rule of masses of vector and spinor fields
 that predicts the mass of Higgs particle $\simeq 310\pm10$ GeV.

\section*{Acknowledgements}

Authors are grateful to A.B. Arbuzov,  D.Yu. Bardin, K.A. Bronnikov,
P. Chankowski, E.A. Kuraev, S. Pokorski, V.B. Priezzhev, A.
Radyushkin, and Yu.P. Rybakov for interesting and critical
discussions.

\end{document}